\documentclass[%
 reprint,
 nofootinbib,
 amsmath,amssymb,
 aps,
]{revtex4-2}

\usepackage{graphicx}
\usepackage{dcolumn}
\usepackage{bm}
\usepackage{textpos}
\usepackage{subfig}
\usepackage{ragged2e}

\begin{document}

\preprint{APS/123-QED}

\title{ MEASUREMENT OF THE HIGGS BRANCHING RATIO BR($H\rightarrow\gamma\gamma$) \\AT 3 TeV CLIC}
\thanks{kacarevicgoran@vin.bg.ac.rs}%

\author{G. Ka\u{c}arevi\'{c}}
\author{I. Bo\v{z}ovi\'{c}-Jelisav\u{c}i\'{c}}%
 \author{N. Vuka\u{s}inovi\'{c}}%
 \author{G. Milutinovi\'{c}-Dumbelovi\'{c}}
  \author{I. Smiljani\'{c}}%
  \author{T. Agatonovi\'{c}-Jovin}%
\affiliation{%
 Vinca Institute of Nuclear Sciences; kacarevicgoran@vin.bg.ac.rs
}%

\collaboration{CLICdp Collaboration}

\author{M. Radulovi\'{c}}
\author{J. Stevanovi\'{c}}
\affiliation{
 University of Kragujevac, Faculty of Science, Radoja Domanovi\'{c}a 12, Kragujevac, Serbia\\
}%

\date{\today}

\begin{abstract}
In this paper we address the potential of a 3 TeV centre-of-mass energy Compact Linear Collider (CLIC) to measure the branching fraction of the Higgs boson decay to two photons, BR($H\rightarrow\gamma\gamma$). Since photons are massless, the Higgs boson coupling to photons is realized through higher order processes involving heavy particles either from the Standard Model or beyond. Any deviation of the measured BR($H\rightarrow\gamma\gamma$), and consequently of the Higgs coupling $g_{H\gamma\gamma}$ from the predictions of the Standard Model, may indicate New Physics. The Higgs decay to two photons is thus an interesting probe of the Higgs sector. 

 This study is performed using simulation of the detector for CLIC and by considering all relevant physics and beam-induced processes in a full reconstruction chain. It is shown that the product of the Higgs production cross-section in $W^+W^-$ fusion and BR($H\rightarrow\gamma\gamma$) can be measured with a relative statistical uncertainty of 5.5\%, assuming the integrated luminosity of 5 ab$^{-1}$ and unpolarized beams. 
\end{abstract}

\maketitle


\section{\label{sec:level1}Introduction}
 
The Higgs boson decay to a pair of photons was one of the discovery channels at the LHC \cite{r1} and also a benchmark process that has shaped requirements for the electromagnetic calorimetry at ATLAS \cite{r2} and CMS \cite{r3}. This channel is also important at proposed $e^{+}e^{-}$ colliders, both in terms of detector performance requirements and complementarity to the expected HL-LHC results \cite{hllhc}. The  combined HL-LHC and future $e^{+}e^{-}$ collider  measurements are expected to give a  statistical uncertainty for the Higgs to photons coupling of $\sim$ 1\% \cite{clic_hllhc}.

\begin{figure}[h]

\centering

\includegraphics[width=6cm]{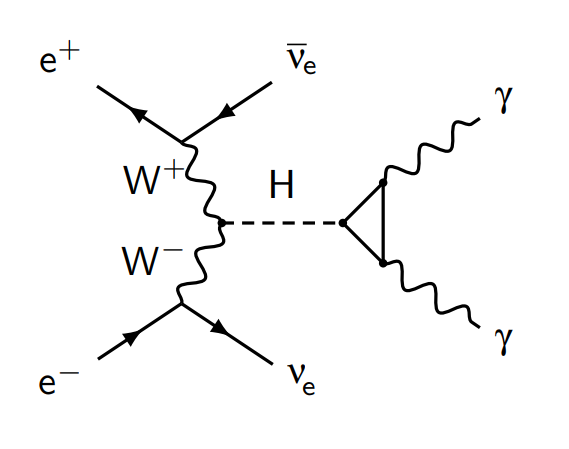} 
\caption{\label{fig-hgg1} Lowest order Feynman diagram of the Higgs production in WW-fusion and subsequent Higgs decay to a pair of photons.}
\end{figure}

 CLIC provides an excellent environment to study the properties of the Higgs boson, including its couplings, with a very high precision. Operation is expected to be staged at three centre-of-mass energies:  at 380 GeV, 1.5 TeV and 3 TeV. WW-fusion (Figure \ref{fig-hgg1}) as the dominant Higgs production mechanism at centre-of-mass energies above $\sim$ 500 GeV will produce large signal yields allowing rare processes such as ${H\rightarrow\mu^+\mu^-}$, ${H\rightarrow Z\gamma}$ and ${H\rightarrow\gamma\gamma}$ to be studied. For a Higgs mass of 126 GeV, the SM prediction for the branching fraction  BR(${H}\rightarrow\gamma\gamma$) is $2.23\times10^{-3}$ \cite{r4}. It is expected that $2\times10^6$ Higgs bosons will be produced at 3 TeV, assuming the nominal integrated luminosity of 5 ab$^{-1}$ which will be used in this paper unless stated otherwise. The signal yield can be increased with the proposed beam polarization by a factor of 1.5 \cite{r5}. The high photon-identification efficiency and good photon energy resolution of a detector for CLIC enable excellent identification of $H\rightarrow \gamma\gamma$ decays.
 
 This paper presents a comprehensive simulation of the experimental measurement of the Higgs production cross-section in WW-fusion $\sigma(e^{+} e^{-} \rightarrow H\nu\bar{\nu}) \times BR(H\rightarrow\gamma\gamma)$ at 3 TeV CLIC. The result of the  study presented here supersedes the estimates based on 1.4 TeV studies given in \cite{r6}. The paper is structured as follows: Simulation and analysis tools are introduced in Section 2, the detector for CLIC is described in  Section 3, while Sections 4 to 6 provide details on signal and background identification and separation, pseudo-experiments and uncertainties of the measurement.

\section{\label{sec:level1}Simulation and Analysis Tools}
The Higgs production in WW-fusion is generated in \break WHIZARD 1.95 \cite{r7}, where a Higgs mass of 126 GeV is assumed. Background processes are also generated in  WHIZARD, using PYTHIA 6.4 \cite{r8} to simulate hadronisation and fragmentation processes. The CLIC luminosity spectrum and beam-induced effects are obtained using GuineaPig 1.4.4 \cite{r9}. Interactions with the detector are simulated using the CLIC\_ILD detector model \cite{r10} within the Mokka simulation package \cite{r11} based on the GEANT4 framework \cite{r12}. Events are reconstructed using the Particle Flow approach (PFA) implemented in the Pandora algorithm \cite{r13}. Photons are reconstructed with PandoraPFA v02-04-00 photon processor \cite{r14}. Simulation, reconstruction and analysis are carried out using  ILCDIRAC \cite{r15}. The TMVA package \cite{r16} is employed for the multivariate analysis classification (MVA) of signal and background events on the basis of their kinematic properties.

\section{\label{sec:level1}Detector for CLIC}
The CLIC\_ILD model is based on the ILD detector proposed for ILC \cite{r17} and it has been modified to the CLIC experimental conditions. The vertex detector is closest to the interaction point to provide reconstruction of secondary vertices for accurate flavor tagging. The Time Projection Chamber is foreseen as the main tracking device providing single point resolution better than 100 $\mu$m in the plane transverse to the beam axis \cite{r10}, together with a low material budget. The CLIC\_ILD detector uses high-granularity electromagnetic (ECAL) and hadronic (HCAL) sampling calorimeters to reconstruct photons and neutral hadrons. The electromagnetic calorimeter is a Silicon-Tungsten calorimeter optimized for longitudinal containment and lateral separation of electromagnetic showers. High granularity in combination with the information from the central tracker leads to an electron identification efficiency of 96\%, while photon identification efficiency is 99\% \cite{r18}. The hadronic calorimeter consists of 60 steel absorbers interleaved with scintillator tiles to contain hadronic showers from neutral hadrons \cite{r10}. A more recent detector model CLICdet \cite{r19} improves the stochastic energy resolution term of the ECAL to 17\% from 20\% of CLIC\_ILD. This, however has no significant impact on the conclusions of this paper \cite{r19}.

\section{\label{sec:level1}Signal and background processes}
\begin{figure} [h]

\centering

\includegraphics[width=6 cm]{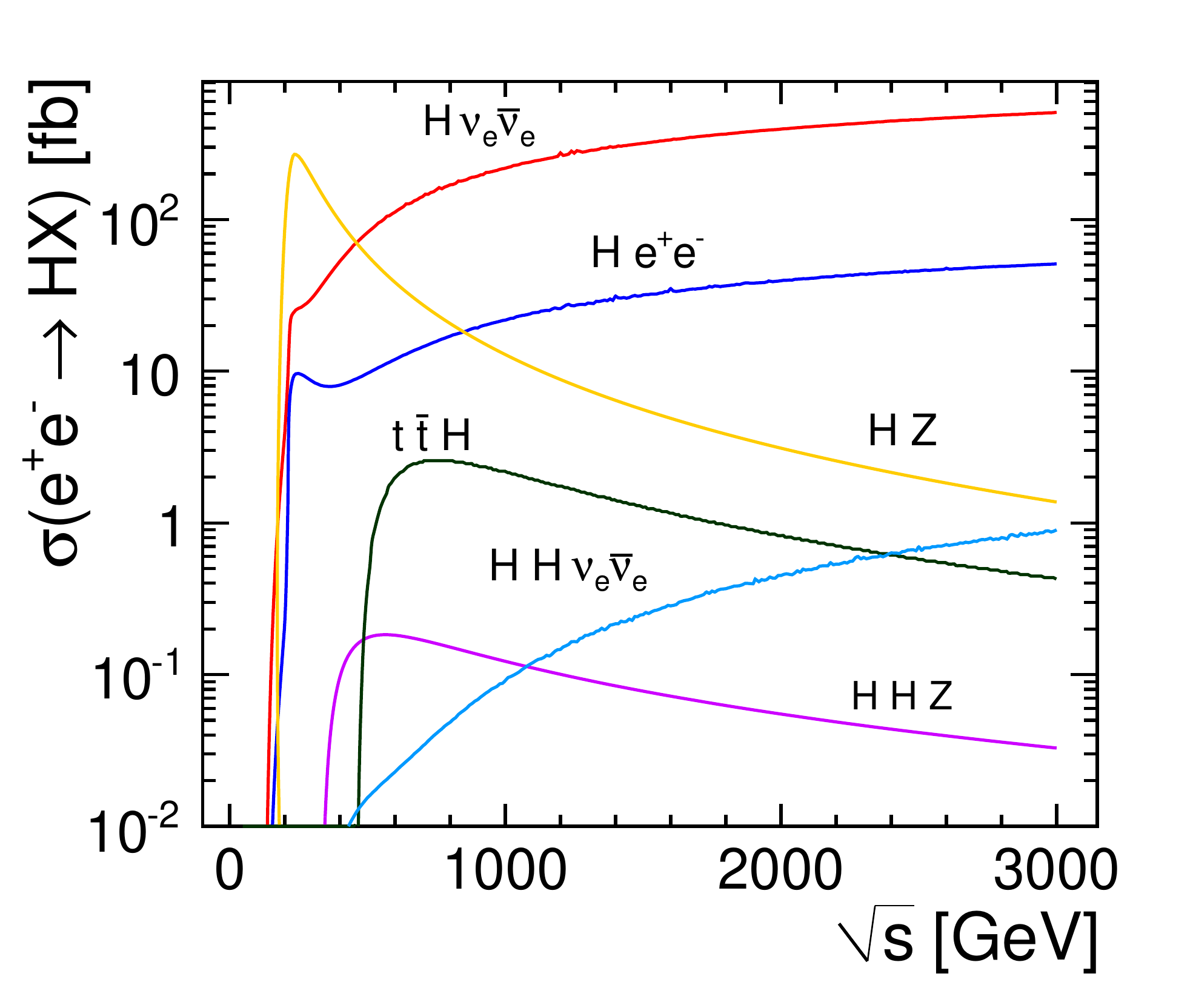}
\caption{\label{fig-xs1}Higgs production cross-sections at different centre-of-mass energies.}

\end{figure}

The main Higgs production processes and backgrounds considered in this paper are summarised in Figure \ref{fig-xs1} and Table \ref{tab-lop}. Higgs boson production at 3 TeV is dominated by the WW-fusion process. Without beam polarization, the effective cross-section for the Higgs production is 415 fb, including Initial State Radiation (ISR) effects as well as a realistic CLIC luminosity spectrum. Taking into account that BR$(H \rightarrow \gamma\gamma)$  is order of 0.23\%, 4750 signal events are expected with the nominal integrated luminosity. In order to describe fully the CLIC experimental environment, simulated Beamstrahlung photons producing hadrons ($\gamma_{BS}\gamma_{BS}\rightarrow hadrons$) are overlaid on each event after the full simulation of the detector response and before the digitization phase. Background processes are considered if two generated photons can be found in the central tracker acceptance  with invariant mass of di-photon system between 100 GeV and 150 GeV. 
Backgrounds arising from mono-photon final states are considered as well if an auxiliary photon (from $\gamma_{BS}\gamma_{BS}\rightarrow hadrons$ overlay, final state radiation or false particle identification) can be found in the detector polar angle acceptance, forming an invariant mass with the final-state photon that falls in the selected window.  

\begin{table}
\caption{\label{tab-lop}Considered signal and background processes with the corresponding effective\footnote{The cross-sections are effective in a sense that condition 100\,GeV $<$ $m_{\gamma\gamma}$ $<$ 150\,GeV is applied to any di-photon system found in the central tracker.} cross-sections \\at 3 TeV centre-of-mass energy. }
\label{parset}
\begin{ruledtabular}
\begin{tabular*}{\columnwidth}{@{\extracolsep{\fill}}llll@{}}
\multicolumn{1}{@{}l}{Signal process}  & $\sigma(fb)$  & N@5 ab$^{-1}$  & N$_{\mathrm{simulated}}$ \\
\hline
$e^+e^-\rightarrow H\nu\nu, H\rightarrow\gamma\gamma$          & 0.95    & 4750  & 24550   \\
\hline
\multicolumn{1}{@{}l}{Background processes}  & $\sigma(fb)$ & N@5 ab$^{-1}$  & N$_{\mathrm{simulated}}$ \\
\hline
$e^+e^-\rightarrow\gamma\gamma$            & 15.2      & $7.6\cdot10^4$  &$3\cdot10^4$   \\ 
$e^+e^-\rightarrow e^+e^-\gamma$           &  335     & $1.7\cdot10^6$ 	& $3\cdot10^6$    \\
$e^+e^-\rightarrow e^+e^-\gamma\gamma$            &  33   & $1.6\cdot10^5$ 	 &$1.5\cdot10^5$          \\
$e^+e^-\rightarrow\nu\bar{\nu}\gamma$           &  13   & $6.6\cdot10^4$	&$2\cdot10^5$             \\
$e^+e^-\rightarrow\nu\bar{\nu}\gamma\gamma$            &  26 & $1.3\cdot10^5$		& $1.6\cdot10^5$         \\
$e^+e^-\rightarrow q\bar{q}\gamma$             & 210     & $1.1\cdot10^6$   & $1.2\cdot10^6$         \\
$e^+e^-\rightarrow q\bar{q}\gamma\gamma$            & 47      &$2.3\cdot10^5$  & $3\cdot10^5$      \\

\end{tabular*}
\end{ruledtabular}

\end{table}


\begin{figure} [h]
 
\centering

\includegraphics[width=0.45\textwidth]{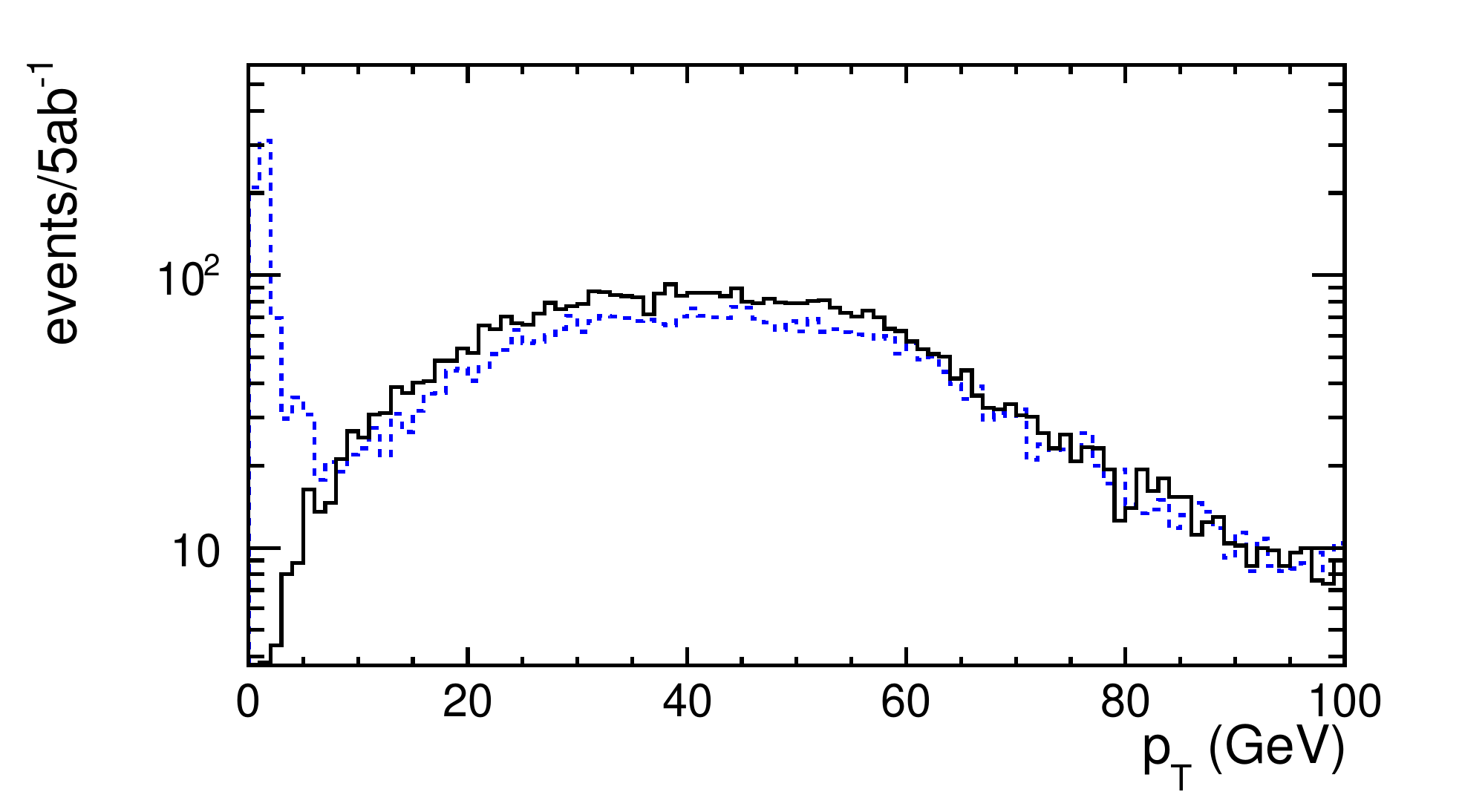}
\begin{textblock}{0.08}(4.6,-2.3)
\bf CLICdp
\end{textblock}
\caption{\label{fig-2nd}The $2^{nd}$ highest reconstructed photons $p_T$  in a signal event (dashed) and the $2^{nd}$ highest $p_T$ photon generated in a Higgs decay (solid). The difference in the two distributions at low $p_T$ values comes from the presence of Beamstrahlung photons at the reconstructed level (dashed).}
 
\end{figure}

\section{\label{sec:level1}Event selection}

\subsection{\label{sec:level2}Photon isolation and Higgs candidate definition}
To ensure that Higgs candidates are found, only events with exactly two isolated photons with transverse momenta greater than 15 GeV are selected. The requirement that both photons have $p_T$ above 15 GeV removes to a great extent photons in a signal event that do not originate from the Higgs decays, as illustrated in Figure \ref{fig-2nd}. We define a photon as isolated if the energy of all reconstructed particles within a 14 mrad cone is less than 20 GeV.  This isolation criterion reduces background processes (in particular $e^+e^-\rightarrow q\bar{q}\gamma$ and $e^+e^-\rightarrow q\bar{q}\gamma\gamma$) by 23\%. Signal loss is negligible. Selection of events with exactly two isolated photons with $p_T > $ 15  GeV results in 22.3\% signal loss, as illustrated  in Figure 4.

\begin{figure} [h]
\centering
\includegraphics[width=0.45\textwidth]{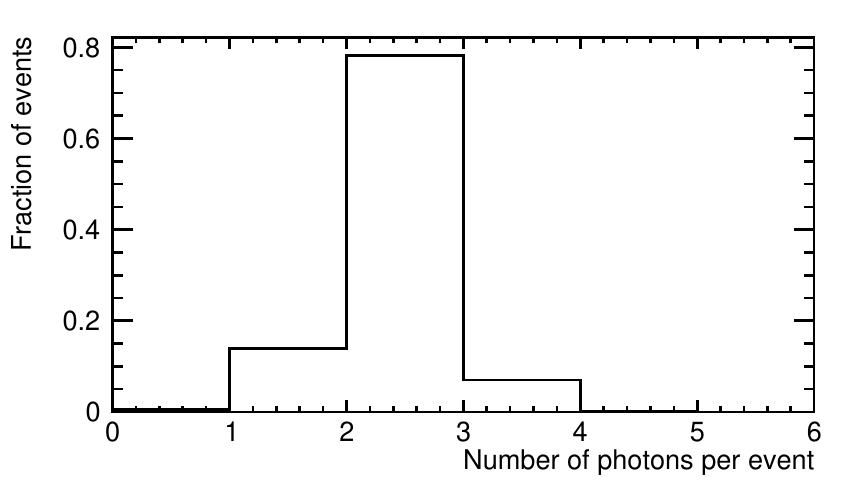}
\begin{textblock}{0.08}(4.6,-2.3)
\bf CLICdp 
\end{textblock}
\caption{\label{fig-ppe1}Number of reconstructed isolated photons per signal event with $p_{T}(\gamma)$ $>$ 15 GeV.}
\end{figure}

\subsection{\label{sec:level2}Preselection}
Signal is separated from backgrounds in a two-stage selection process: preselection and MVA based selection. The preselection suppresses high cross-section \break backgrounds like $e^{+}e^{-} \rightarrow e^{+}e^{-}\gamma$ and $e^{+}e^{-} \rightarrow e^{+}e^{-}\gamma\gamma$. Preselection variables are optimized as follows:
\begin{itemize}
\item Reconstructed di-photon invariant mass in the range from 110 GeV to 140 GeV, corresponding to the Higgs mass window,
\item Reconstructed di-photon energy in the range between 100 GeV and 1000 GeV,
\item Reconstructed di-photon transverse momentum in the range between 20 GeV and 600 GeV.

\end{itemize}

\begin{figure}[h]%
    \centering

    \subfloat[]{\label{fig-energy}{\includegraphics[width=0.45\textwidth, height=31ex]{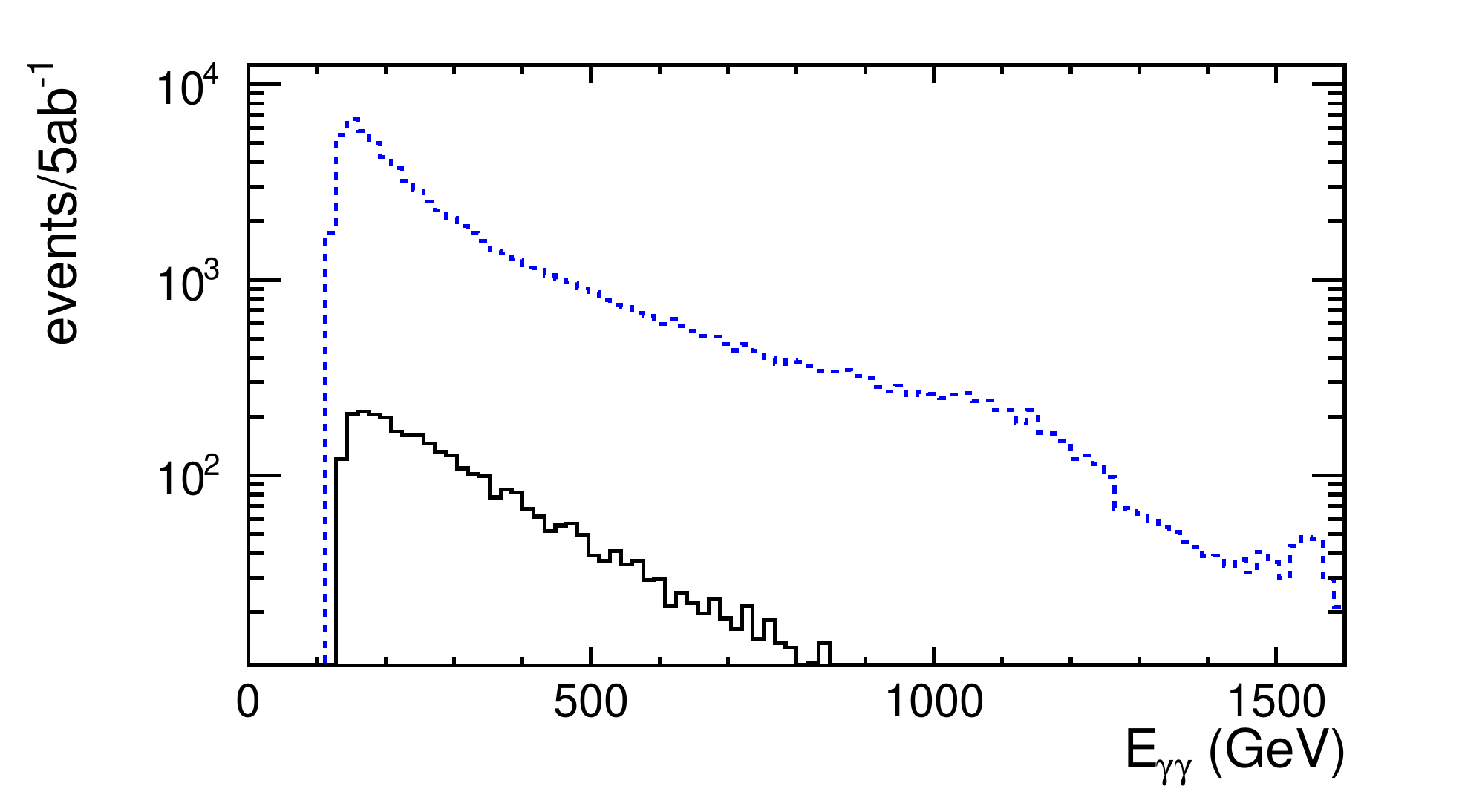} }}
    \begin{textblock}{0.08}(4.6,-2.35)
    \bf CLICdp 
    \end{textblock}
    \qquad
    \subfloat[]{\label{fig-pt}{\includegraphics[width=0.45\textwidth, height=31ex ]{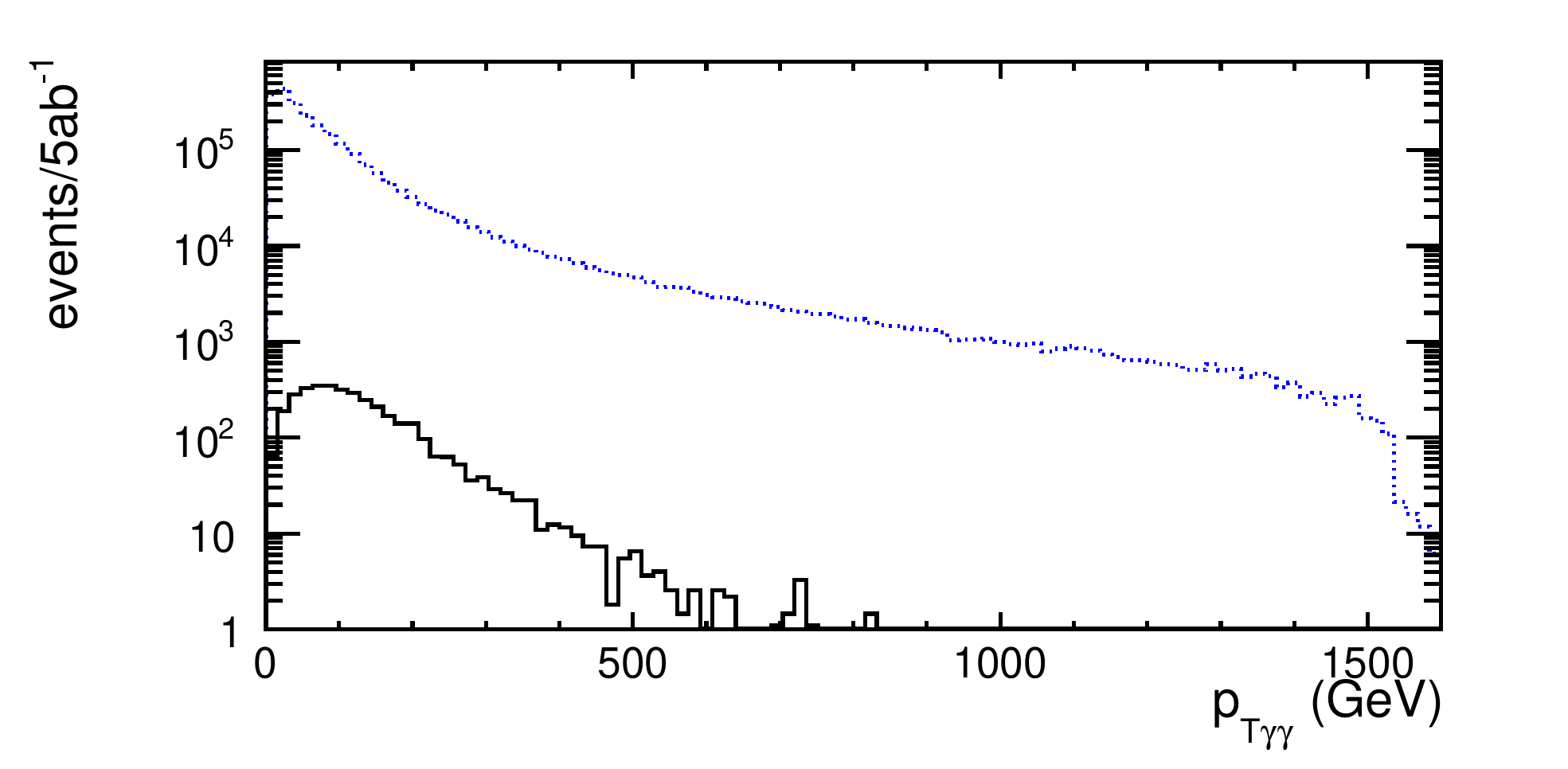} }}
        \begin{textblock}{0.08}(4.6,-2.35)
        \bf CLICdp 
        \end{textblock}%
    \label{fig-23}%
    \caption{Higgs candidate observables for signal and background: energy (a) and transverse momentum (b). Signal is represented with the solid line while background is represented as dashed.}
\end{figure}

\begin{figure}[h]
 
\centering

\includegraphics[width=7.2cm, height = 6cm]{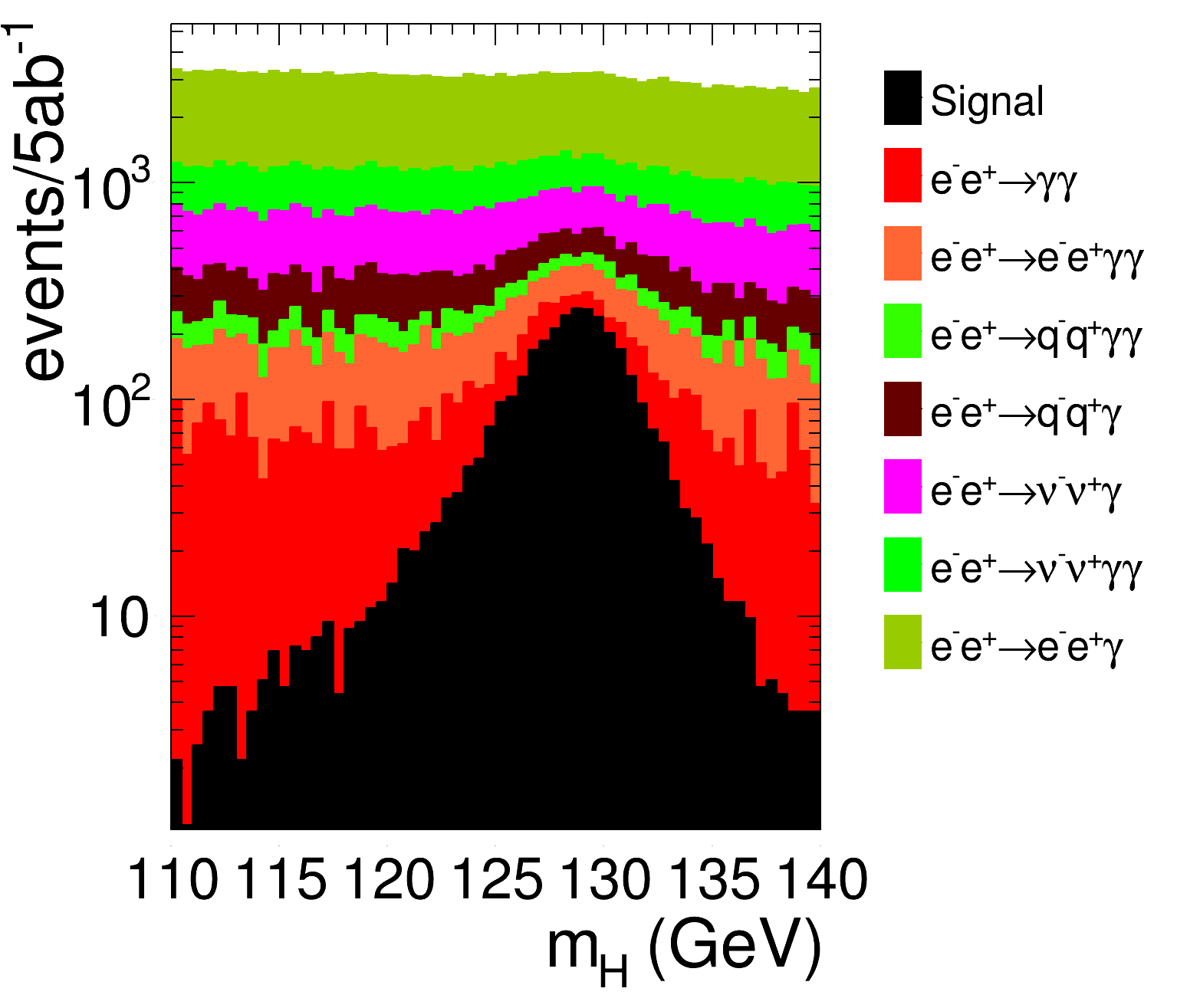}
    \begin{textblock}{0.08}(3.2,-3.1)
    \bf CLICdp 
    \end{textblock}
\caption{\label{fig-preslection}Stacked histograms of Higgs mass distributions for signal and background after preselection.}
\end{figure}

Distributions of di-photon energy and transverse momentum are given in Figure \ref{fig-energy} and Figure \ref{fig-pt} respectively, illustrating the choice of selection range. The signal and background di-photon invariant mass after preselection is given in Figure \ref{fig-preslection}. Preselection efficiency for signal is 70\% and background dominates over the signal by a factor of 25.

\subsection{\label{sec:level2}Multivariate analysis}

\begin{figure}[h]

\centering

\includegraphics[width=7.5cm, height=4.5cm]{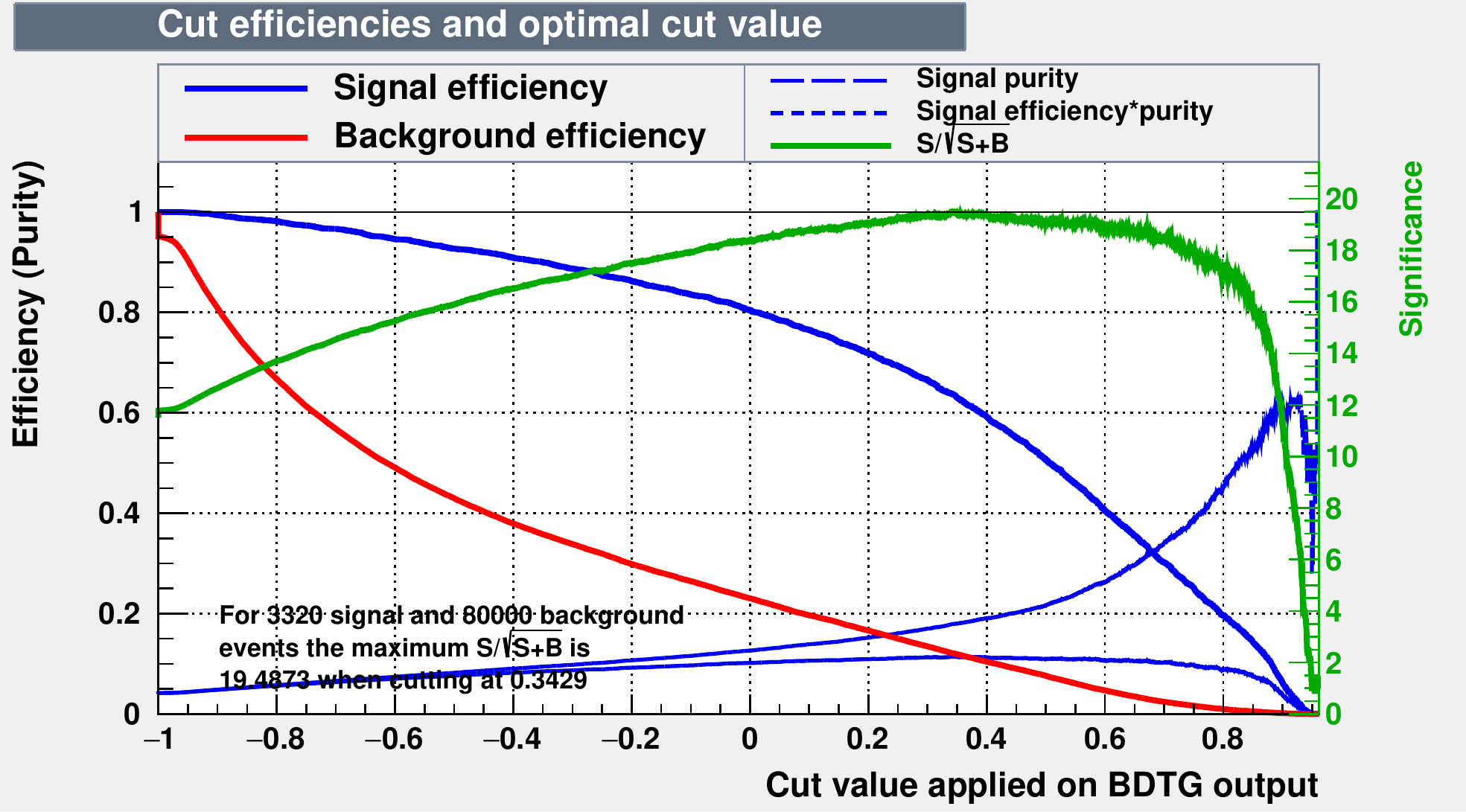}
    \begin{textblock}{0.08}(5.,-2.36)
    \bf CLICdp 
    \end{textblock}
\caption{\label{fig-BDTG}BDTG performance in the training phase.}

\end{figure}

Preselected signal and background events are further separated using an MVA method based on the Gradient Boosted Decision Trees (BDTG). Twelve observables are used for classification of events: di-photon energy, di-photon transverse momentum, di-photon polar angle, cosine of the helicity angle,  transverse momenta of photons, polar angle of photons,  energy of photons, total ECAL energy per event and total HCAL energy per event.  The optimal cut-off value of the BDTG output variable was found to be 0.34, as illustrated in Figure \ref{fig-BDTG}. Variables are sufficiently uncorrelated for MVA to perform optimally.

\begin{figure}

\centering

\includegraphics[width=7.5cm, height=6.2cm]{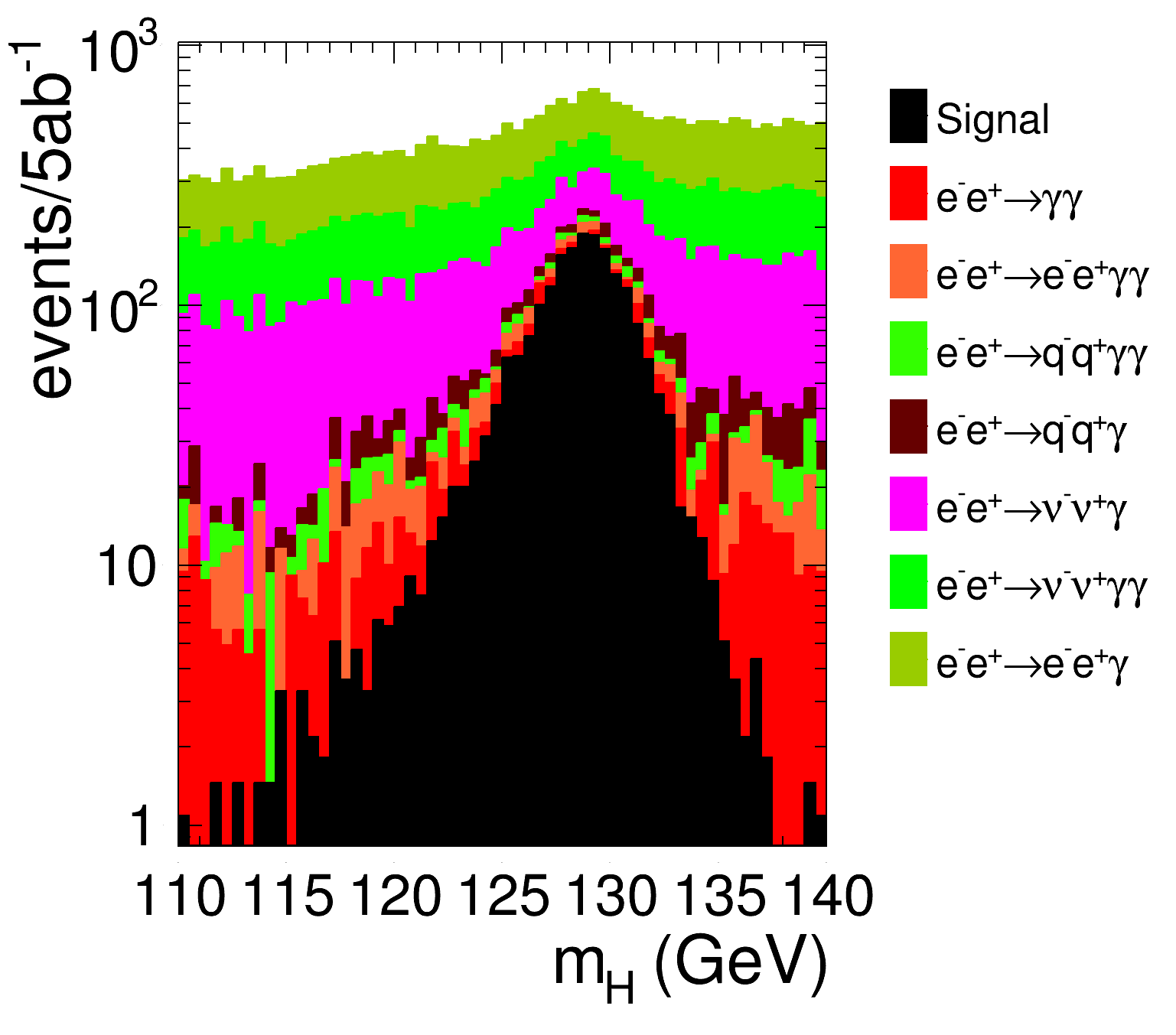}
\begin{textblock}{0.08}(3.3,-3.12)
\bf CLICdp
\end{textblock}
\caption{\label{fig-stackMVA}Stacked histograms of Higgs mass distributions for signal and background after MVA selection.}

\end{figure}

\noindent The classifier cut was selected to maximize statistical significance defined as:
\begin{equation}
    S  =  \frac{N_{s}}{\sqrt{N_{s}+N_b{}}} \label{eq-signigicance}
\end{equation}
 where $N_{s}$ and $N_{b}$ are number of  signal and background events after the MVA selection. The MVA efficiency for  signal is 62.7\%, resulting in an overall signal selection efficiency of 43.7\%, corresponding to a signal yield of 2080 selected Higgs candidates. The remaining background after the MVA application is $\sim$ 10 times larger than the signal and originates mostly from the processes such as ${e^+e^-\rightarrow\nu\bar{\nu}\gamma}$ and \break ${e^+e^-\rightarrow \nu\bar{\nu}\gamma\gamma}$ or from a high cross-section process like\break ${e^+e^-\rightarrow e^+e^-\gamma}$. The Higgs candidate mass distribution after MVA selection is illustrated in Figure \ref{fig-stackMVA}, giving the composition of the background.

\section{\label{sec:level1}Pseudo experiments}

\subsection{\label{sec:level2}Method}

The observable to be measured is a product of the Higgs production cross-section and a corresponding branching fraction for Higgs di-photon decay and it can be experimentally determined from the counted number of signal events $N_{s}$ as:
\begin{equation}
    \sigma(e^+ e^- \rightarrow H\nu\bar{\nu}) \times BR (H\rightarrow \gamma \gamma)  =  \frac{N_{s}}{L \cdot \epsilon_{s} } \label{eq-observable}
\end{equation}

 \noindent where $L$ represents the integrated luminosity, $\epsilon_{s}$ is the overall signal efficiency including detector acceptance, photon identification efficiency and signal selection efficiency. The number of signal events will be determined from combined fit of di-photon invariant mass distributions of selected simulated (or experimental) data with the function $f$:
 \begin{equation}
f (m_{\gamma\gamma}) = N_{s}\cdot f_{s}(m_{\gamma\gamma})  +  N_{b}\cdot f_{b} (m_{\gamma\gamma}) \label{eq-pdf}
 \end{equation} 
 where $N_{s}$ and $N_{b}$ are number of selected signal and background events, and $f_{s}$ and $f_{b}$ are the probability density functions (PDF) describing $m_{\gamma\gamma}$ for signal and background respectively. These PDFs are determined from  simulated samples of signal and background data.

\subsection{\label{sec:level2}Signal and background PDF}
Functions $f_{s}$ and $f_{b}$ from Eq. \ref{eq-pdf} are used to fit the fully simulated datasets of signal and background after the  the MVA selection phase. The signal PDF consists of two Gaussian functions, one describing the tail ($f_{flat}$) and the other describing exponential part ($f_{exp}$) of di-photon mass distribution of the signal:
\begin{eqnarray}
\label{eq-pfds}
f_s &=& f_{flat} + C_{1} \cdot f_{exp}  \\
f_{flat} &=& \left\{ \begin{array}{rl}\nonumber
		e^{-\frac{ (m_{\gamma\gamma} - m_{H})^2 }
          {2 \sigma^2 + \beta_L (m_{\gamma\gamma} - m_{H})^2 } }
	&    \hspace{3mm}  	,(m_{\gamma\gamma} < m_{H}) \\
		e^{-\frac{ (m_{\gamma\gamma} - m_{H})^2 }
          {2 \sigma^2 + \beta_R (m_{\gamma\gamma} - m_{H})^2 } } 
	& 	   \hspace{3mm} ,(m_{\gamma\gamma} > m_{H})
	\end{array} \right. \label{eq-pdf-signal} \\
f_{exp} &=& \left\{ \begin{array}{rl}
		e^{-\frac{ (m_{\gamma\gamma} - m_{H})^2 }
          {2 \sigma^2 + \alpha_L |m_{\gamma\gamma} - m_{H}| } }
	&     \hspace{3mm}	,(m_{\gamma\gamma} < m_{H}) \\
		e^{-\frac{ (m_{\gamma\gamma} - m_{H})^2 }
          {2 \sigma^2 + \alpha_R |m_{\gamma\gamma} - m_{H}| } } 
	&  \hspace{3mm} ,(m_{\gamma\gamma} > m_{H}),
	\end{array} \right. \nonumber  
\end{eqnarray}

\begin{figure}

\centering
\includegraphics[width=0.35\textwidth]{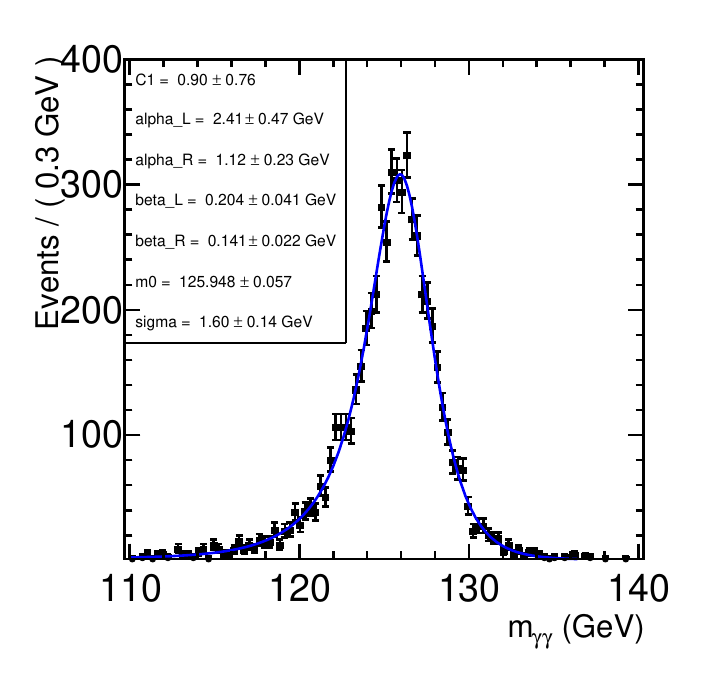}
\begin{textblock}{0.08}(4.1,-3.1)
\bf CLICdp
\end{textblock}
\caption{\label{fig-signalFit} Fit of di-photon invariant mass of the selected signal (points) and the fit function $f_{s}$ (line).}

\end{figure}

\noindent where $\sigma, C_{1}, \alpha_{L,R}, \beta_{L,R}$, as well as Higgs mass $m_{H}$ are free parameters determined by the fit (Figure \ref{fig-signalFit}). The fit is performed using RooFit \cite{r20}.

\begin{figure}
  	
\centering
\includegraphics[width=0.4\textwidth]{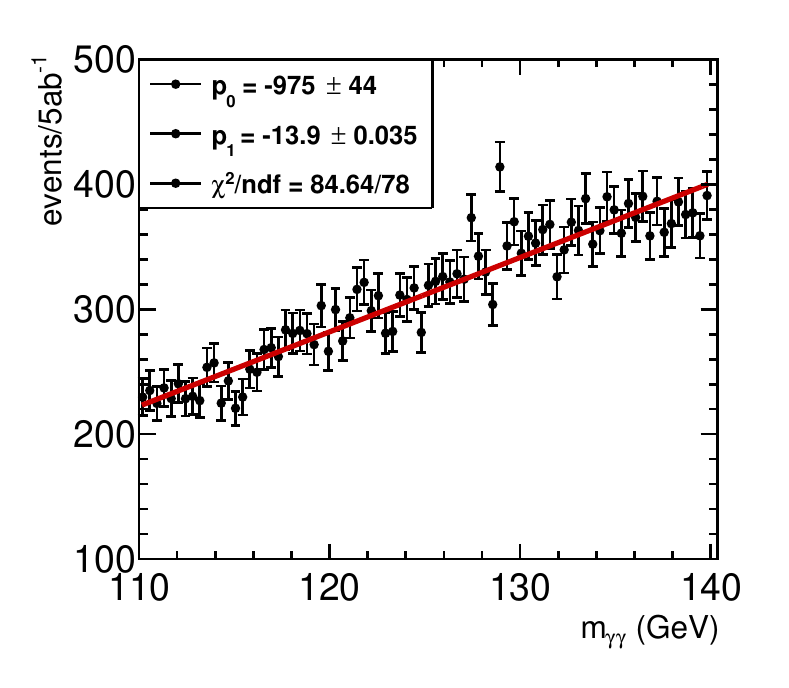}
\begin{textblock}{0.08}(4.3,-3.17)
\bf CLICdp
\end{textblock}
\caption{\label{fig-background} Di-photon invariant mass $m_{\gamma\gamma}$ for the sum of all background processes remaining after event selection (points). The fit function given in Equation 5 is overlaid (line). }
\end{figure}

The di-photon mass distribution for background is fitted with a linear function $f_{b}$: 
\begin{equation}
f_{b} = p_{0} + p_{1}\cdot m_{\gamma\gamma} \label{eq-bkd}
\end{equation}

\noindent where $p_{0}$ and $ p_{1}$ are free parameters of the fit. The fit of background di-photon invariant mass distribution is illustrated in Figure \ref{fig-background}, and shows no sensitivity to the SM Higgs mass.

\begin{figure}[h]

\centering

\includegraphics[width=6.9 cm]{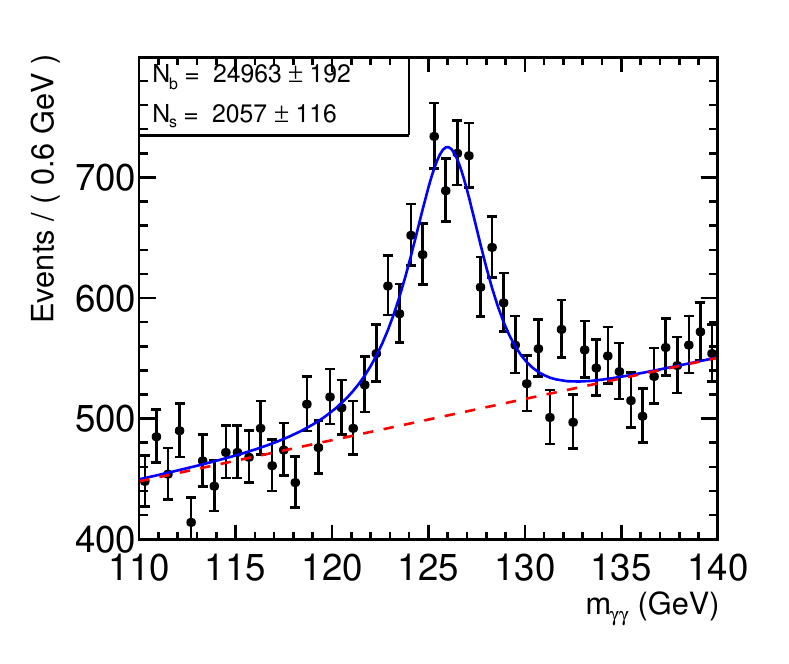}
\begin{textblock}{0.08}(4.3,-2.95)
\bf CLICdp
\end{textblock}
\caption{\label{fig-fitexample} Example of one pseudo-experiment, showing di-photon invariant mass of pseudo-data (black), corresponding fit with the function $f$ from Eq. 3 (full line) and background fit  with function $f_{b}$ (dashed line)  from Eq. 5.}

\end{figure}

\subsection{\label{sec:level2}Pseudo-experiments}
\begin{figure} [h]
 
\centering

\includegraphics[width=6.9cm]{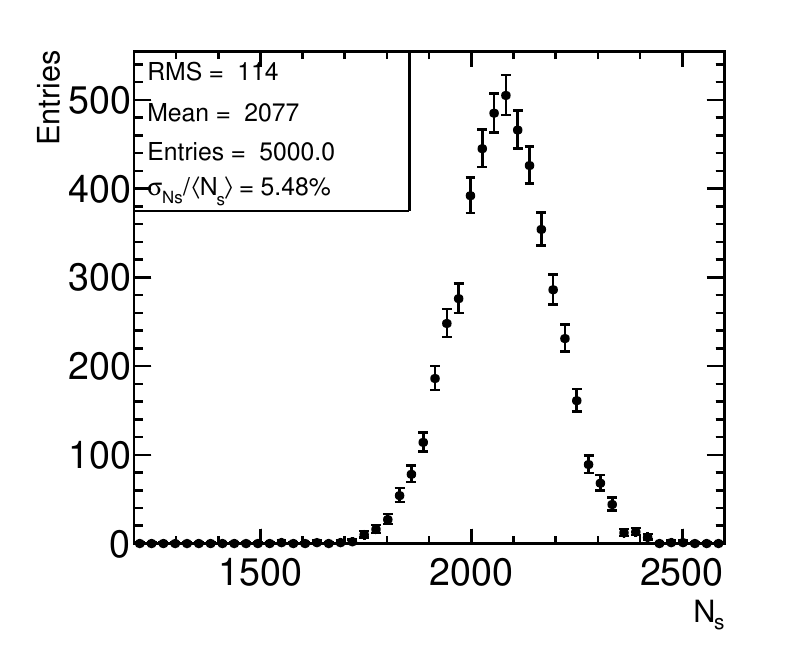}
\begin{textblock}{0.08}(4.3,-2.99)
\bf CLICdp
\end{textblock}
\caption{\label{fig-toyMC}Pull distribution of 5000 pseudo-experiments.}
\end{figure}

The pseudo-data distribution, combining both signal and background after MVA selection, is fitted with function $f$ (Eq. \ref{eq-pdf}), where $N_{s}$ and $N_{b}$ are set as free parameters. In this way the number of signal events is determined in the same way it would be on a set of experimental data. Such a measurement we call a pseudo-experiment. An example of one pseudo-experiment is shown in Figure \ref{fig-fitexample}. In order to estimate the statistical dissipation of the measured number of signal events, 5000 pseudo-experiments with 5 ab$^{-1}$ of data were performed. Pseudo-data for signal is randomly picked from fully simulated signal sample, while $m_{\gamma\gamma} $ distribution for background is generated from background PDF by randomly changing parameters $p_{0}$ and $p_{1}$ from Eq.5. The RMS of the resulting pull distribution  over all pseudo-experiments is taken as the estimate of the statistical uncertainty of the measurement (Figure \ref{fig-toyMC}). It reads that the statistical uncertainty of the extracted number of signal events is 5.5\%. 

\subsection{\label{sec:level2}Systematic uncertainty}

Several sources of systematic uncertainty of the measurement are considered. The systematic uncertainty associated with photon identification requires more detailed investigation as it depends on the distribution of material and details of the treatment of converted photons within the particle flow algorithm. As converted photons account for of order 10\% of all photons, the overall systematic uncertainty from this source is expected to be smaller than the statistical uncertainty. Assuming for illustration an uncertainty on the photon identification efficiency of 0.5\%, it  would result in a systematic uncertainty of about 1\% on the BR$(H \rightarrow \gamma\gamma)$ measurement. The relative uncertainty of the integrated luminosity, and hence of the measured cross-section, is expected  to be of order of several permille at CLIC \cite{ilclumi}. Another source of systematic uncertainty is due to uncertainty of the luminosity spectrum recontruction. In \cite{r21} it has been shown that the CLIC luminosity spectrum at 3 TeV centre-of-mass energy can be corrected better than 5\% above 50\% of the nominal centre-of-mass energy, while above 75\% of the nominal centre-of-mass energy the corresponding uncertainty  of the  correction is at a permille level \cite{r22}. As discussed in \cite{r6}, the impact of uncertainty of the luminosity spectrum reconstruction on $H\nu\bar{\nu}$ production at 3 TeV (in $H \rightarrow b\bar{b}$ channel) is found to be of order of several permille. The energy resolution of the ECAL also has the permille-level impact on preselection efficiency. If we assume the relative uncertainty of the ECAL sampling term of 10\%\footnote{As shown in \cite{r23} and \cite{r24} ECAL can be usually calibrated with the relative uncertainty of the sampling term  $\sim$ 10\%.}, resulting uncertainty of reconstructed photon energy of $\sim$ \space40 MeV has a negligible effect on $N_{s}$ determination. Similarly, the uncertainty of di-photon transverse momentum as a preselection variable hardly contributes to the systematic uncertainty of the measurement. To probe systematic sensitivity if the result to background modeling, linear fit from the Equation \ref{eq-bkd} was replaced with the second order polynomial function. Negligible (permile) level impact is found. With the considerations above, relative systematic uncertainty of the measurement is expected to be smaller than the statistical one. 

\section{\label{sec:level1}Summary}

The accessibility of  WW-fusion as a dominant Higgs production mechanism at energies of 500 GeV and above enable the Higgs rare  decays at 3 TeV CLIC to be measured. Excellent performance of the electromagnetic calorimeter  to identify high-energy photons together with the overall PFA reconstruction of physics processes enables the measurement of the loop induced Higgs decays to two photons at the percent level. In the full simulation of experimental measurement,  we have shown that $\sigma(e^{+} e^{-} \rightarrow H\nu\bar{\nu})\times BR(H\rightarrow\gamma\gamma)$  can be measured  at 3 TeV CLIC with a relative statistical uncertainty of 5.5\%, assuming 5 ab$^{-1}$ of integrated luminosity and unpolarized beams. This result can be further improved with the proposed beam polarization scheme, which would increase the Higgs production cross-section by a factor of 1.5, due to the chiral nature of WW-fusion as a charged-current interaction. The systematic uncertainty is estimated to be smaller than the statistical one. This analysis completes the set of Higgs to $\gamma\gamma$ measurements foreseen at CLIC energy stages above 1 TeV centre-of-mass energy.

\begin{acknowledgments}
The work presented in this paper has  been carried out in the framework of the CLIC detector and physics study (CLICdp) collaboration and the authors would
like to thank the CLICdp members for their support.  We are in particular grateful to the colleagues from the Analysis Working Group for useful discussions, to Philipp Roloff, for useful ideas exchanged in the course of the analysis and to Aleksander Filip Zarnecki who contributed to the paper by raising several important questions. Our special thanks goes to Nigel Watson for editing the text. We acknowledge the support received until 2020 from the Ministry of Education, Science and Technological Development of the Republic of Serbia within the national project OI171012.
\end{acknowledgments}

\end{document}